\documentclass[conference]{IEEEtran}
\IEEEoverridecommandlockouts
\usepackage{cite}
\usepackage{amsmath,amssymb,amsfonts}
\usepackage{algorithmic}
\usepackage{graphicx}
\usepackage{textcomp}
\usepackage{xcolor}
\def\BibTeX{{\rm B\kern-.05em{\sc i\kern-.025em b}\kern-.08em
    T\kern-.1667em\lower.7ex\hbox{E}\kern-.125emX}}
\usepackage{hyperref}       
\usepackage{listings}
\usepackage{url}
\usepackage{pgf,tikz,pgfplots}

\usepackage{subcaption}    
\pgfplotsset{compat=1.14}
\begin{document}

\title{Don't Persist All: Efficient Persistent Data Structures}

\author{\IEEEauthorblockN{Pratyush Mahapatra}
\IEEEauthorblockA{Computer Sciences\\
University of Wisconsin Madison\\
Madison, WI 53703 \\
\texttt{pratyush@cs.wisc.edu}}
\and
\IEEEauthorblockN{Mark D. Hill}
\IEEEauthorblockA{Computer Sciences\\
University of Wisconsin Madison\\
Madison, WI 53703 \\
\texttt{markhill@cs.wisc.edu}}
\and
\IEEEauthorblockN{Michael M. Swift}
\IEEEauthorblockA{Computer Sciences\\
University of Wisconsin Madison\\
Madison, WI 53703 \\
\texttt{swift@cs.wisc.edu}}}

\maketitle

\begin{abstract}
Data structures used in software development have inbuilt redundancy to improve software reliability and to speed up performance \cite{Taylor:1980:RDS:1313350.1313933}. Examples include a Doubly Linked List which allows a faster deletion due to the presence of the previous pointer. With the introduction of Persistent Memory, storing the redundant data fields into persistent memory adds a significant write overhead, and reduces performance. In this work, we focus on three data structures - Doubly Linked List, B+Tree and Hashmap, and showcase alternate partly persistent implementations where we only store a limited set of data fields to persistent memory. After a crash/restart, we use the persistent data fields to recreate the data structures along with the redundant data fields. We compare our implementation with the base implementation and show that we achieve speedups around 5-20\% for some data structures, and up to 165\% for a flush-dominated data structure.     
\end{abstract}


\section{Introduction}

With Intel's Optane DC Persistent Memory \cite{intel} commercially available, we can safely say Persistent Memory technology is finally here. It presents a huge shift from the current memory organization where memory is divided into two distinct categories - DRAM for main memory which allows byte-level access and low latency but is volatile and has low density, and storage on flash or HDD which is extremely dense and persistent but is orders of magnitude slower \cite{262588213843476} and where memory accesses are limited to block granularity. The introduction of NVM allows us to get the best of both worlds. It provides persistence with byte-level accesses at near-DRAM speeds and with a much higher density. For example the largest DRAM DIMM available is from Samsung and has a capacity of 128 GB, whereas Intel's largest Optane DC PMMs has a capacity of 512 GB \cite{intel}. It's also much cheaper than DRAM with a 128 GB Intel Optane DIMM costing around \$600 whereas the DRAM of the same capacity costs as much as \$4500 \cite{alcorn_2019}. 

With a lower price/byte ratio, one potential use for this new technology is as a DRAM replacement, and using it as a vast memory while ignoring persistence. However, in order to make the most of this new technology and use the opportunity to gain high speed persistence, there are many hurdles that we have to overcome \cite{rudoff2017persistent}. For example, with the memory being persistent, ensuring data consistency becomes vital. In particular, if the system crashes while an update is being made to a data structure in Persistent Memory, the data structure may be left in a corrupted state as the update is only half-done. To maintain consistency, we need to prevent reordering of writes to Persistent Memory by the CPU and the Memory Controller while also manually flushing the persistent data structures using cache-flush and fence instructions which incur significant performance overhead.

In this paper, we focus on improving the performance of the Data Structures stored in Persistent Memory by reducing the number of flushes needed. We do this by not persisting redundant data fields from data structures, which are put in place for added software reliability and performance \cite{Taylor:1980:RDS:1313350.1313933}, and storing consistently the essential data fields in Persistent Memory. For example, only storing the "NEXT" pointer in a Doubly Linked List, and recreating the "PREVIOUS" pointer after a crash. We call our approach Partly Persistent since only some of the data fields are made persistent. Our approach produces an average speed up of \texttt{15\%} over a naive implementation which persists all data fields. We also showcase our reconstruction algorithms which allow rebuilding of all the redundant data fields in volatile memory using only the persistent data fields.   

The rest of the paper is organized as follows. Section 2 gives an overview of related work in the field. In Section 3, we discuss our motivation for this work. Section 4 talks about the Design and Implementation of the Partly Persistent Data Structures. It also showcases the Reconstruction Algorithms for each data structure. We show the Performance Evaluation in Section 5 along with some of the insights that we have gained from our experiments. Section 6 is a Discussion Section and mentions future work that could be conducted in this field, and we conclude in Section 7.

\section{Related Work}
There have been numerous papers on using Persistent Memory, and there will be many more as we try to solve the challenges facing it. In this section, we discuss some of the most relevant work. One way of utilizing Persistent Memory is through the use of transactions which ensures data consistency and atomicity. Mnemosyne \cite{Volos:2011:MLP:2248487.1950379} provides a simple programming interface to Programmers through the use of lightweight transaction mechanism. NV-Heaps \cite{coburn2012nv} is another way to manage Persistent Memory directly by using an persistent object system based on epochs. Both Mnemosyne and NV-Heaps provides an interface for managing Persistent Memory, however they did not delve into the problem of optimizing implementations for Persistent Memory which we do in this work. NOVA FS \cite{194454} is a file system created for systems with Persistent Memory. NOVA FS has Radix Trees in Volatile Memory to speed access to logged data but it does not store the tree persistently. This is similar to the approach we take in our work, where we avoid storing redundant data in Persistent Memory.   
HOPS in WHISPER \cite{Nalli:2017:APM:3093315.3037730} introduces the idea of a lightweight fence called ofence for ordering and using a heavier primitive called dfence to ensure durability. HOPS looks at reducing the overhead of ensuring consistency, and proposes two fence primitives. This work can use HOPS dfence/ofence to further the overhead of consistency among the essential data fields that need to be made persistent. There has been work on adapting data structures for Persistent Memory as well. NVC-Hashmap \cite{schwalb2015nvc} uses a Split-Ordered List implementation for a Hashmap implementation since the layout of the structure allows atomicity control. It does not look into the impact of flushing data. The most relevant work to this paper is NVTree \cite{yang2015nv} which uses similar concepts of critical data and reconstructable data to create a B+Tree design optimized for Persistent Memory.    
\section{Motivation}
\subsection{Cost of Persistence}
\begin{figure}[h]
    \centering
\framebox{
\begin{tikzpicture}
\begin{axis}[
    ybar,
    xlabel={Number of Flushes},
    ylabel={Execution Time in seconds},
    xtick={125000000,250000000,375000000,500000000},
    xticklabels={62.5M,125M,250M,500M},
    ytick={40, 60,80,100,120,140, 160,180,200,220},
    yticklabels={40, 60, 80, 100,120,140, 160,180,200,220},
    legend pos=north west,
    ymajorgrids=true,
    grid style=dashed,
]
\addplot
    coordinates {
 (500000000,219.6391)
(375000000, 141.8069)
(250000000,101.3296)
(125000000,77.55254)
    };

\end{axis}
\end{tikzpicture}
} 
\caption{Cost of Persistence}
\label{fig:costpersistence}
\end{figure}
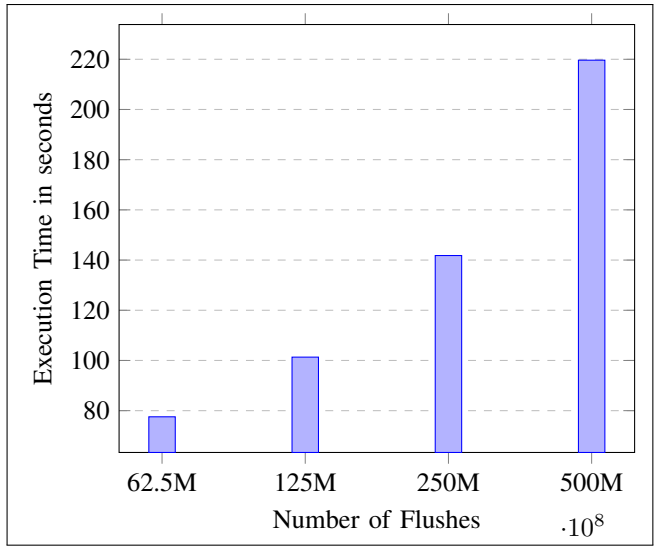

The addition of Persistent Memory to the fray brings forward the problem of ensuring data consistency to the limelight. Ensuring Data Consistency implies that the programmer must explicitly flush the data and add memory fences to avoid data being reordered by the CPU and the Memory Controller. Unfortunately, the cost of such operations is high and hurts the performance of the system.\\ 
To get a quantitative estimate of the cost of flushing, we ran a micro-benchmark of single-threaded Append-only operations on a simple Linked List structure. In this benchmark, depending on the user-input, either all the nodes or only some of the nodes were flushed out. This test was run on a 1 GHz Intel Machine with Optane DCPMM \cite{intel}. More details of this test machine can be found in Section \ref{testmachine}. Figure \ref{fig:costpersistence} shows the result from this test run. We find that there is almost a linear increase in execution time as we increase the number of cache line flushes in the system. \\
One potential way of reducing this cost would be by collating multiple persistent operations together, but that becomes tricky when trying to ensure data consistency. We wanted to look at the possibility of reducing the number of cache line flushes and find the minimum number required to ensure that data is not lost. 

\subsection{Redundancy in Data Structure Designs}
Widely used data structures commonly have inbuilt redundancies in them \cite{Taylor:1980:RDS:1313350.1313933}. This could be for purposes of ensuring software reliability or improving the performance of the system. For example, a Doubly Linked List has both a previous pointer as well as a next pointer. This allows the linked list to be traversed in both directions, thus improving performance while also improving software reliability. Similarly, B+Trees have nodes that do not contain any data but are used to create a tree structure which allows efficient traversal and fast access to the data nodes. These nodes can be recreated by using information from the leaf nodes. \\
These data structures are extremely useful and improve system performance but using them, as they are, in Persistent Memory would lead to more time spent in memory operations. For example, storing the redundant data fields which are useful for traversal in volatile memory like both the PREV and NEXT pointers in a Doubly Linked List could potentially double the number of flushes and thus increase the program execution time.    
\section{Partly Persistent Data Structure Design and Implementation}
Storing only the absolute minimum number of data fields and using them to recreate the redundancy would help in reducing both the number of flushes and also ensure that the performance of the data structure is maintained while using it in volatile memory.
For our work, we chose data structures that are commonly used and which we expect would be used with Persistent Memory applications in the near future. We went with the following three data structures for this study:
\begin{itemize}
    \item Doubly Linked List : Common data structure used in many applications like Scheduler, LRU Cache, Playlists, etc.
    \item B+Trees : Data structure that's commonly used in Databases and File Systems \cite{comer1979ubiquitous}  
    \item Hashmap : Key-Value structure data structure used in many applications.  
\end{itemize}
\subsection{Design Decisions}
Our purpose was to identify the minimum number of data fields that need to be made persistent by flushing their values to Persistent Memory and not lose any data. We also wanted to be able to create an efficient reconstruction algorithm which recreates the data structure by using only the persistent data fields. 
\subsection{Using Persistent Memory}
We used Intel's Optane DCPMM for all our experiments. We used this machine in the App Direct Mode \cite{intel} which allows applications to write directly to Persistent Memory using byte-level accesses. To ensure we only wrote to Persistent Memory, we created a file and mmaped it to the Persistent Memory with the MAP\_SYNC and MAP\_SHARED\_VALIDATE flags set. We then used pointer arithmetic to locate the data structures in the file allocated in Persistent Memory.  
\subsection{Doubly Linked List}
\subsubsection{Design}
We started off with the design of the simplest data structure among the selected ones, a doubly linked list. The Linked List structure has three members- "DATA", "NEXT" and "PREV", where NEXT points to the succeeding node in the list and PREV points to the preceding node in the list. DATA is designed as a \texttt{struct} of integer values. For our optimized implementation, only the NEXT and DATA are stored persistently and their values are flushed out. Listing \ref{lst:ll} shows the implemented Node structure. The data structure supports two write operations - append and delete, both of which are constant time operations.
\begin{minipage}{0.49\textwidth}
\begin{lstlisting}[frame=single, language=C, caption=Linked List Node Structure, commentstyle=\color{red}, label={lst:ll}]
typedef struct Value Value;
struct Value{
    long long value; 
    long long padding1; 
    long long padding2; 
    long long padding3; 
    long long padding4; 
    long long padding5; 
    long long padding6; 
};
struct Node{
    Value data;          
    //Persistent Data Field 
    struct Node *next;   
    //Persistent Data Field
    //64B Cache Line Boundary
    struct Node *prev;
}__attribute__((__aligned__(64)));
\end{lstlisting}
\end{minipage}

\begin{figure*}[!t]
  \centering
  \includegraphics[width=\textwidth]{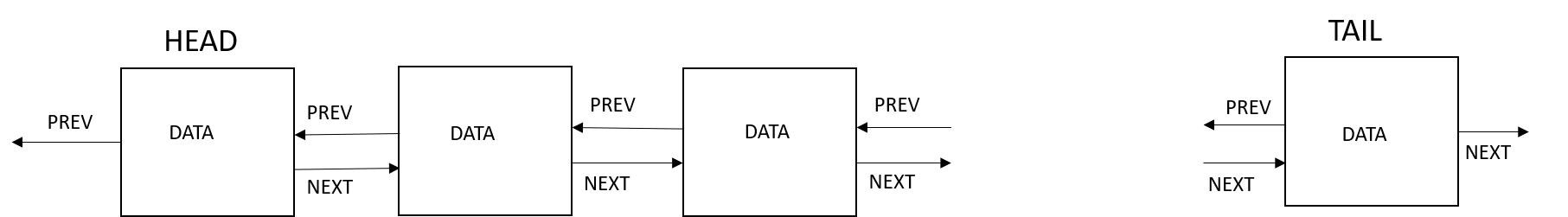}
  \caption{Visual Representation of our Linked List Design}
  \label{fig:llnode}
\end{figure*}

\subsubsection{Implementation}
We have two distinct implementations of the data structure. In the first implementation, which we refer to as Fully Persistent, we treat all data fields of the data structure as persistent and stored all of them in persistent memory. In the second implementation, which we call Partly Persistent, we only treat a subset of the data fields as persistent and only flush those members to persistent memory. We also have two distinct versions for the Partly Persistent implementation. The first implementation is used for performance evaluation since we wanted to avoid the overhead of calling malloc (which we do in our second implementation) to impact our results. The second implementation focuses on reliability and crash correctness. In the first version, we operate directly on the Persistent Memory. Thus all the data structure operations impact the data stored in persistent memory. In the second version, the data structure operates on volatile memory and is mapped to Persistent Memory only after a flush operation. Thus the flush operation acts as a checkpoint for the Persistent Memory. Any corruption that happens before a flush is done is not reflected in Persistent Memory. For example, if the NEXT pointer points to itself in volatile memory and that portion of the data is not flushed out, it will not affect the Persistent data structure.        
\subsubsection{Reconstruction Algorithm}
On a crash, we run a Reconstruction Algorithm when the program starts to recreate the lost data structure. We use the DATA and NEXT pointers stored in Persistent Memory to reconstruct the Doubly Linked List structure. To reconstruct the Data Structure, we first read the head of the file where the data structure is stored. It also stores a flag bit which indicates if the Persistent Linked List was safely initialized. If the flag is set, we point to the top of file, which has a NODE stored which points us to the HEAD of the Linked List. If the data structure is not corrupted, we move to the HEAD node of the Linked List. From here the replication is straight forward since we can traverse the Linked List using the NEXT pointers which have been stored persistently and also restore the PREV pointers on the way. On the completion of one complete forward traversal, we are able to restore all the PREV pointers in all the nodes and also store the address of the TAIL node.  
\subsection{B+Trees}
\subsubsection{Design}
Next we tackled a more complex data structure, the B+Tree. We took an existing design of a B+Tree \cite{interactive_b_tree} and modified it for our purpose. The B+Tree was implemented using the \texttt{struct} "node" which has the following data fields - "POINTERS", "KEYS", "PARENT", "IS\_LEAF", "NUM\_KEYS" and "NEXT". If the node is a LEAF, then the POINTERS point to a "record" \texttt{struct} which stores a \texttt{struct} of integers called "VALUE". For our optimized implementation, we only store the LEAF nodes persistently here, which reduces the number of nodes stored by a factor of $(1 - 1/n)*(t/(t - 1))$ where \texttt{n} is the total number of LEAF nodes and \texttt{t} is the average occupancy per intermediate nodes. We can then use the reconstruction algorithm to recreate the tree. Listing \ref{lst:btree} shows the code for the Node structure as it is implemented. The data structures supports - "INSERT", "DELETE" and "FIND" operations. 
\begin{minipage}{0.49\textwidth}
\begin{lstlisting}[frame=single, language=C, caption=B+Tree Node Structure, commentstyle=\color{red}, label={lst:btree}]

#define DEFAULT_ORDER 19

typedef struct Value Value;
struct Value{
    long long value; 
    long long padding1; 
    long long padding2; 
    long long padding3; 
    long long padding4; 
    long long padding5; 
    long long padding6; 
    long long padding7; 
}; //64B
typedef struct record {
	Value value;
} record;

typedef struct node node;
struct node {
	void * pointers[DEFAULT_ORDER]; 
	//152B
	int  keys[DEFAULT_ORDER - 1]; 
	//72B
	struct node * parent; //8B
	bool is_leaf; //1B 
	//Only nodes where is_leaf=TRUE 
	//is Persistent
	int num_keys; //4B
	struct node * next; //8B 
}__attribute__((__aligned__(64))); 
//256B (Four cachelines)

\end{lstlisting}
\end{minipage}

\begin{figure*}[!t]
  \centering
  \includegraphics[width=\textwidth]{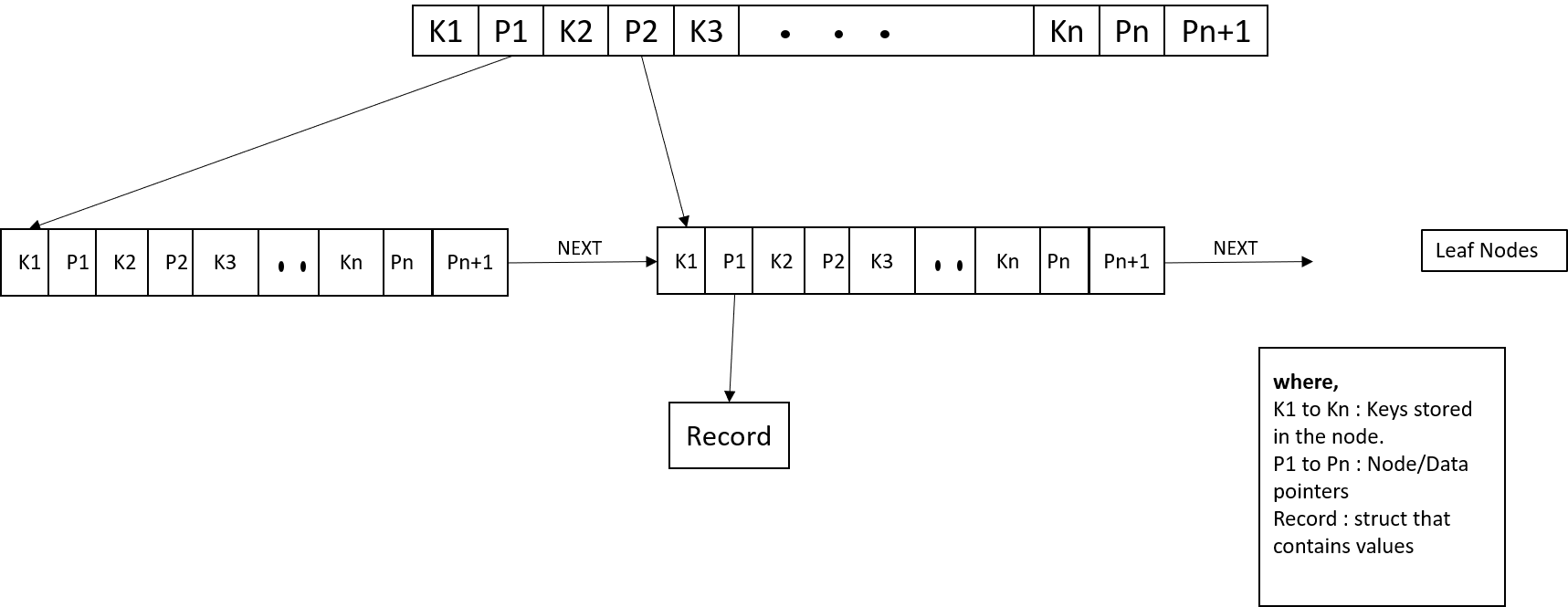}
  \caption{Visual Representation of our B+Tree design}
  \label{fig:b+tree}
\end{figure*}

\subsubsection{Implementation}
The B+Tree is implemented as a tree of NODES where each NODE points to multiple CHILD nodes. Each LEAF node (a node which does not have any CHILDREN nodes) points to multiple "record" \texttt{struct}s which contains the VALUE stored. The tree is traversed by comparing the KEY in the input query with the KEY values present in the node. The array of KEYS in the node is traversed till a KEY larger than the input KEY is found and then we traverse to the CHILD node pointed to by that KEY, or create the CHILD node if it is not present (for INSERTS). If none of the KEYs present are larger than the node KEYS, we traverse to the right-most CHILD node pointed to in the node. Hence, there is always one more CHILD node pointer than the number of KEYS present. There is also a constraint with respect to the number of CHILD nodes each node can point to. Each NODE can point to a maximum of $X$ CHILD nodes and should have a minimum of $X/2$ CHILD nodes. If these properties are not met, then there is a re-balancing of the tree which could entail deleting some nodes or transferring nodes from one PARENT to other. \\ 
For the purpose of our project, we have two distinct implementations. Similar to Linked List, we have a Fully Persistent Implementation and a Partly Persistent Implementation. In the partly persistent implementation, only the LEAF nodes and their corresponding "record" \texttt{struct}s are stored persistently. This implies on an INSERT or a DELETE operation, only the LEAF node is flushed out. Whereas, in the Fully Persistent Implementation, all the nodes are stored persistently, and thus all the nodes where values are changed or updated, have to be flushed out. Again similar to the Linked List, we have two implementations of the Partly Persistent implementation, one for performance studies and the second for correctness. As noted previously, flushes act as checkpoints for correctness.
\subsubsection{Reconstruction Algorithm}
After a crash, on running the program again, we use the Reconstruction Algorithm to recreate the data structure.
We use the LEAF nodes to reconstruct the B+Trees. As earlier, we first point to the top of the file which contains a node. If the value present in the node is valid, which can be found out by reading the "NUM\_KEYS" data field in the node, then the B+Tree has been safely initialized and we can move forward with our reconstruction. The value of the "POINTERS[0]" in the node points to the left most CHILD node (node with the smallest KEYS). In the next step, we find all the CHILD nodes of the B+Tree. This is possible since the LEAF nodes of B+Tree form a linked list, and hence all of the nodes can be found by traversing the Linked List using the "NEXT" pointer. Once, we have all the CHILD nodes, we can then arrange them in buckets of specific size with each bucket being assigned a single intermediate node. This step continues recursively till there is only one bucket left and a single HEAD node is allocated. \\ 
Choosing the size of the BUCKET is a design decision that can have an impact on the performance of the data structure. 
\begin{itemize}
    \item Setting the bucket size to the maximum permissible value which is equal to the total number of nodes a single node can point to, helps reduce the number of intermediate nodes created and the number of levels in the B+Tree. It also speeds up "FIND" and "DELETE" operations since the tree traversal encounters lesser number of intermediate levels.
    \item Setting the bucket size to the minimum permissible value (maximum/2 + 1), leads to larger number of intermediate nodes and larger number of levels, but it reduces the number of tree rebalancing operations needed due to "INSERTS" and thus speeds up "INSERT" operations.
\end{itemize}
In our evaluation, we set the bucket size to the maximum permissible value which is set to 19 in our evaluation which allows the node to match the inherent granularity of DCPMM - 256 bytes.
\subsection{Hashmap}
\subsubsection{Design}
For our last data structure, we tackled Hashmap, a data structure that's commonly used in key-value store applications for its constant time lookup operations. We took an existing Hashmap implementation from the Android Open Source Project directory\cite{google_git} and modified it for our purpose. The data structure is implemented using \texttt{struct} Hashmap which has the following data fields - "BUCKETS", "BUCKETCOUNT", "HASH FUNCTION", "EQUALS FUNCTION", "LOCK" and "SIZE". "BUCKETS" data field points to an array of \texttt{struct} Entry which has "KEY", "VALUE", "HASH" and "NEXT" data fields. For our optimized implementation, we only store "SIZE" data field from \texttt{struct} Hashmap and the "KEY" and "VALUE" data fields from \texttt{struct} Entry. "BUCKETS" and "BUCKETCOUNT" can be recreated through the "SIZE" data field whereas "HASH" can be recreated using "KEY" and "VALUE" data fields. Listing \ref{lst:hm} shows the Hashmap Node Structure and also makes note of the data fields that are stored persistently. The data structure supports "INSERT", "REMOVE" and "FIND" operations. 

\begin{minipage}{0.49\textwidth}
\begin{lstlisting}[frame=single, language=C, caption=Hashmap Node Structure, commentstyle=\color{red}, label={lst:hm}]
typedef struct Value Value;
struct Value{
    long long value1; 
    long long value2; 
    long long value3; 
    long long value4; 
    long long value5; 
    long long value6; 
    long long value7; 
}; //56B

typedef struct Entry Entry;
struct Entry {
    long long key; //Persistent Data Field
    Value value;  //Persistent Data Field
    //64B Cache Line Boundary
    int hash;
    Entry* next;
}__attribute__((__aligned__(64)));

struct Hashmap {
    size_t bucketCount;
    int (*hash)(int key);
    bool (*equals)(int keyA, int keyB);
    mutex_t lock; 
    size_t size;  //Persistent Data Field
    Entry** buckets;
}__attribute__((__aligned__(64)));
\end{lstlisting}
\end{minipage}

\begin{figure*}[!t]
  \centering
  \includegraphics[width=\textwidth]{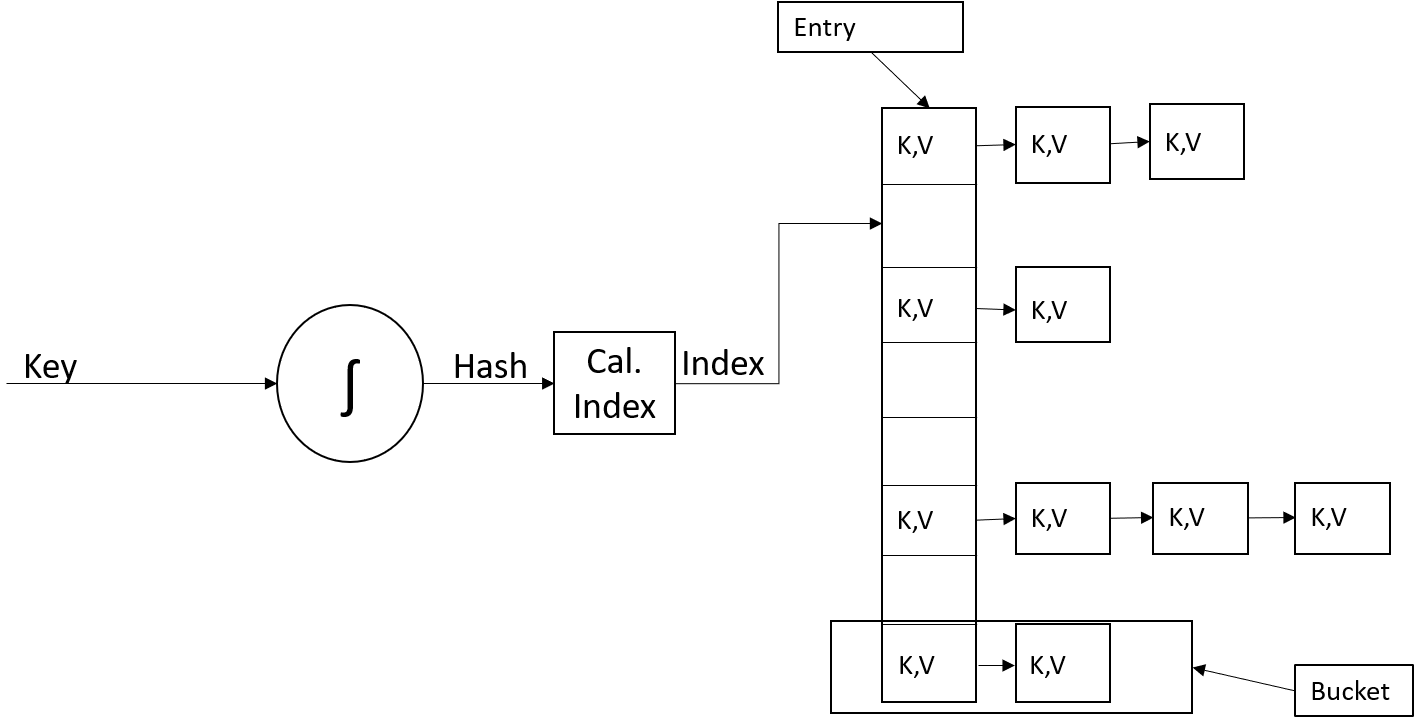}
  \caption{Visual Representation of our Hashmap design}
  \label{fig:hashmap}
\end{figure*}
\subsubsection{Implementation}
Hashmap is implemented as an array of BUCKETS with each BUCKET being represented as a linked list of ENTRIES. Each ENTRY stores a KEY and the VALUE as well as the HASH value and a pointer to the NEXT entry. On an INSERT, the input KEY is sent to the HASH function which outputs a HASH value on which a modulus operation is conducted to ensure that the outputted value points to a valid BUCKET. The corresponding value is then used to map to the entry in the BUCKET array. If the entry in the BUCKET array is NULL, the element is inserted at that position. Else the ENTRY is traversed using the NEXT pointer until the end of the list. DELETE and FIND operations are implemented similarly. \\
For the purpose of our project, we have two distinct implementations. Similar to the two data structures described above, we have a Fully Persistent and Partly Persistent implementations. In the Partly Persistent Implementations only a subset of the data fields are made persistent, i.e. SIZE data field in the \texttt{struct} Hashmap and KEY, VALUE in \texttt{struct} Entry. The Fully Persistent Implementation on the other hand has all data fields stored persistently. Following the trend of the other two data structures, we also have two different implementations of the Partly Persistent data structure, one for performance studies and the second for correctness. Here too, flushes act as checkpoints for correctness.
\subsubsection{Reconstruction Algorithm}
After a crash, we use the reconstruction algorithm on running the program to recreate the lost data structure and find the entries. We used the spatial arrangement of the \texttt{struct} Entry and the SIZE data field to recreate the Hashmap. We first read the file which has stored the \texttt{struct} Hashmap to check if the Hashmap is valid and then read the SIZE data field. The SIZE data field is used to recover BUCKETCOUNT data field. Then, we read the file which stores the \texttt{struct} Entry. Since the \texttt{struct} Entry are stored adjacently in the same file, we can use information from the SIZE data field and travel through all the valid entries stored in the file. If the node has a KEY which is \textbf{not} NULL, it is a valid Entry and we use the KEY value to recover the HASH value. The HASH value is then used to calculate the Index where the Entry is stored in the BUCKETS array. If the BUCKET is empty, we store it in the first position, otherwise we traverse the list till pointer points to NULL and store our Entry at that position. We then set NEXT data field of the Entry to NULL. We continue doing this till we have found all the Entries which is denoted by SIZE and thus recreate the Hashmap.    
\section{Evaluation} 
For our evaluation, we focus on performance gains while ensuring correctness. We use the naive Fully Persistent Implementation of all three data structures as the base and compare it to the Partly Persistent Implementation. We use workloads that do either Inserts/Updates or Deletes and we do not look at Find/Search operations since they do not update the data structure and thus do not have any dirty data to be written back to Persistent Memory. In all the variations of our workloads shown below, we conduct a fixed number of write operations (inserts + deletes) which is set to 100M with the data structure initialized with 200M entries.  
\subsection{Experimental Setup} \label{testmachine}
We have early access to a machine equipped with Intel Optane DCPMM (aka Real Persistent Memory). The Machine is a 96-core Intel Cascade Lake running at a frequency of 1 GHz. The machine has been configured to run in 100\% App Direct Mode \cite{ilkbahar_ilkbahar_2018} which allows byte-addressable access of Persistent Memory to the applications. \\ 
For our experiments, we use a B+Tree Maximum Node Capacity of \texttt{19} which makes our node traverse four cache-lines (256B) and a default Hashmap Load Factor of \texttt{0.75}.
\subsection{Insertion Performance}
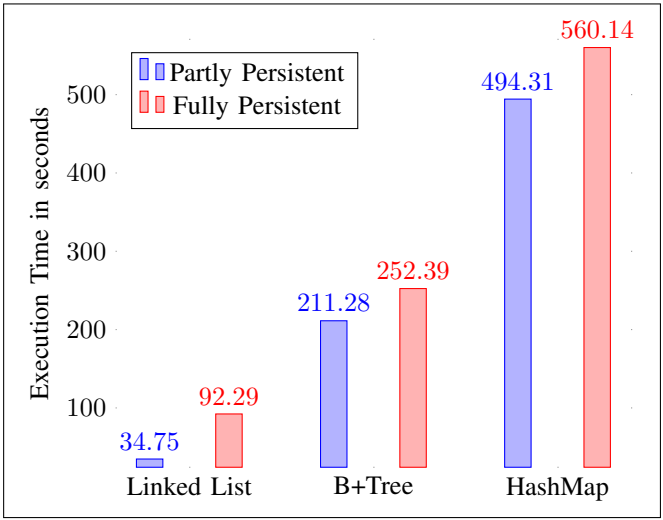
\begin{figure}[h]
    \centering
\framebox{
\begin{tikzpicture}
  \begin{axis}[
    ybar = 20pt,
    x axis line style = { opacity = 0 },
    ylabel            = {Execution Time in seconds},
    tickwidth         = 0pt,
    enlarge x limits  = 0.2,
    enlarge y limits  = 0.02,
    symbolic x coords = {Linked List, B+Tree, HashMap},
    xtick = data,
    nodes near coords,
    legend pos=north west
  ]
\addplot
    coordinates {
(Linked List, 34.749) (B+Tree, 211.279) (HashMap, 494.306)
    };
\addplot
    coordinates {
(Linked List, 92.293) (B+Tree, 252.394) (HashMap, 560.136)
    };
\legend{Partly Persistent, Fully Persistent}

\end{axis}
\end{tikzpicture}
} 
\caption{Insert-only Workload : Execution Time}
\label{fig:append}
\end{figure}

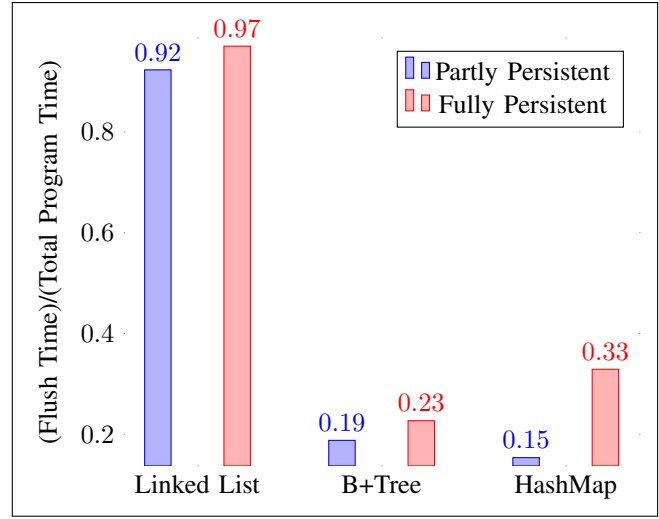
\begin{figure}[h]
    \centering
\framebox{
\begin{tikzpicture}
  \begin{axis}[
    ybar = 20pt,
    x axis line style = { opacity = 0 },
    ylabel            = {(Flush Time)/(Total Program Time)},
    tickwidth         = 0pt,
    enlarge x limits  = 0.2,
    enlarge y limits  = 0.02,
    symbolic x coords = {Linked List, B+Tree, HashMap},
    xtick = data,
    nodes near coords,
    legend pos=north east
  ]
\addplot
    coordinates {
(Linked List, 0.923) (B+Tree, 0.188) (HashMap, 0.1537)
    };
\addplot
    coordinates {
(Linked List, 0.97) (B+Tree, 0.2269) (HashMap, 0.329)
    };
\legend{Partly Persistent, Fully Persistent}

\end{axis}
\end{tikzpicture}
} 
\caption{Insert-only Workload : Proportion of Time Spent in Flush Operations over Total Execution Time}
\label{fig:appendflushproportion}
\end{figure}

In this experiment, we use a single-threaded workload which only inserts new elements into each data structure. To avoid page translation overheads, we create a translation for all the pages mmaped by reading each page once. Figure \ref{fig:append} has the results for all 3 data structures and Figure \ref{fig:appendflushproportion} shows the amount of time each data structure spends in the flush operation in proportion to its total execution time. We find that for the Linked List data structure, where Insert is a constant-time operation, flush takes the majority of time. And thus by cutting down on time spent in flush operations, we are able to achieve speedups of \texttt{165\%}. For the B+Tree and Hashmap data structures, where flush time is not the dominant factor, we are still able to manage speedups of \texttt{19.45\%} and \texttt{13.3\%} respectively.

\subsection{Deletion Performance}
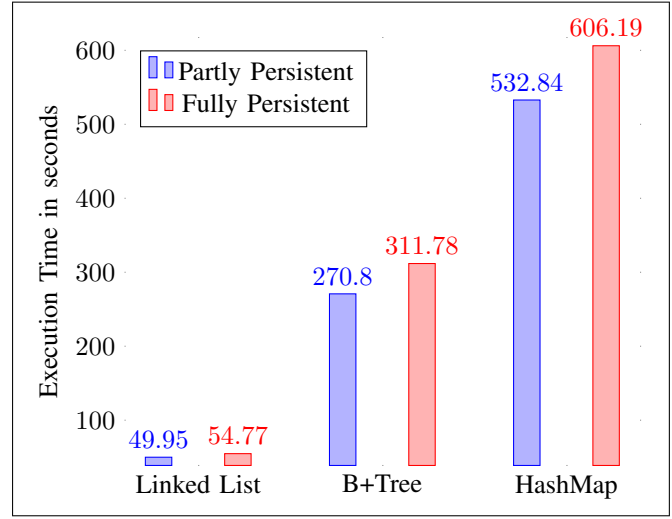
\begin{figure}[h]
    \centering
\framebox{
\begin{tikzpicture}
  \begin{axis}[
    ybar=20pt,
    x axis line style = { opacity = 0 },
    ylabel            = {Execution Time in seconds},
    tickwidth         = 0pt,
    enlarge x limits  = 0.2,
    enlarge y limits  = 0.02,
    symbolic x coords = {Linked List, B+Tree, HashMap},
    xtick = data,
    nodes near coords,
    legend pos=north west
  ]
\addplot
    coordinates {
(Linked List, 49.949) (B+Tree, 270.797) (HashMap, 532.837)
    };
\addplot
    coordinates {
(Linked List, 54.772) (B+Tree, 311.779) (HashMap, 606.193)
    };
\legend{Partly Persistent, Fully Persistent}

\end{axis}
\end{tikzpicture}
}     
\caption{Delete-only Workload : Execution Time}
\label{fig:delete}
\end{figure}

In this experiment, we run a delete-only workload over a data structure after initializing it with multiple entries. Figure \ref{fig:delete} has the execution time results of all 3 data structures and Figure \ref{fig:deleteflushproportion} shows the proportion of time each data structure spends in dealing with flushes in proportion to the entire program execution time. Focusing on the Linked List data structure, we find that the data structure still spends a lot more time on flushing. The amount of flushes in a delete operation is lesser than the time taken in an insert operation, and thus we get a smaller speedup of \texttt{9.6\%}. For B+Tree and Hashmap, we get a speedup of \texttt{15.13\%} and \texttt{13.76\%} respectively.     

\pgfkeys{/pgf/number format/.cd,fixed,precision=2}
\begin{figure}[h]
    \centering
\framebox{
\begin{tikzpicture}
  \begin{axis}[
    ybar = 20pt,
    x axis line style = { opacity = 0 },
    ylabel            = {(Flush Time)/(Total Program Time)},
    tickwidth         = 0pt,
    enlarge x limits  = 0.2,
    enlarge y limits  = 0.02,
    symbolic x coords = {Linked List, B+Tree, HashMap},
    xtick = data,
    nodes near coords,
    legend pos=north east
  ]
\addplot
    coordinates {
(Linked List, 0.9648) (B+Tree, 0.0655) (HashMap, 0.1484)
    };
\addplot
    coordinates {
(Linked List, 0.9579) (B+Tree, 0.1309) (HashMap, 0.3068)
    };
\legend{Partly Persistent, Fully Persistent}

\end{axis}
\end{tikzpicture}
} 
\caption{Delete-only Workload : Proportion of Time Spent in Flush Operations over Total Execution Time}
\label{fig:deleteflushproportion}
\end{figure}
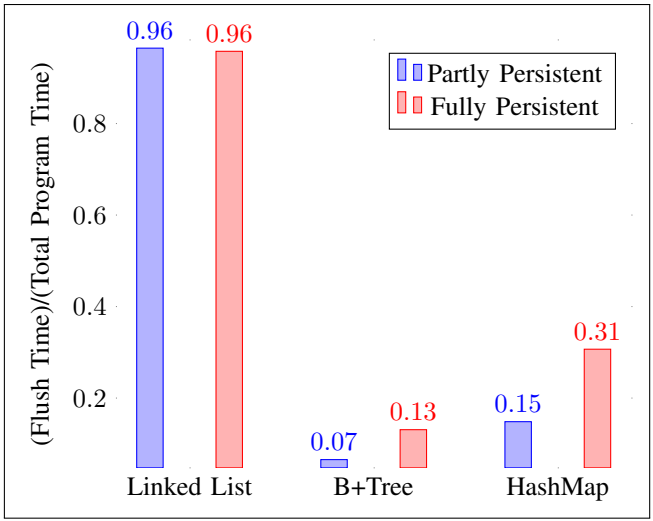

\subsection{Mixed Workloads}
\begin{figure}[h]
    \centering
\framebox{
\begin{tikzpicture}
  \begin{axis}[
    ybar=20pt,
    x axis line style = { opacity = 0 },
    ylabel            = {Execution Time in seconds},
    tickwidth         = 0pt,
    enlarge x limits  = 0.2,
    enlarge y limits  = 0.02,
    symbolic x coords = {Linked List, B+Tree, HashMap},
    xtick = data,
    nodes near coords,
    legend pos=north west
  ]
\addplot
    coordinates {
(Linked List, 53.753) (B+Tree, 242.961) (HashMap, 589.240)
    };
\addplot
    coordinates {
(Linked List, 80.535) (B+Tree, 267.202) (HashMap, 715.617)
    };
\legend{Partly Persistent, Fully Persistent}

\end{axis}
\end{tikzpicture}
} 
\caption{4:1 (Insert/Delete) Workload : Execution Time}
\label{fig:4-1}
\end{figure}
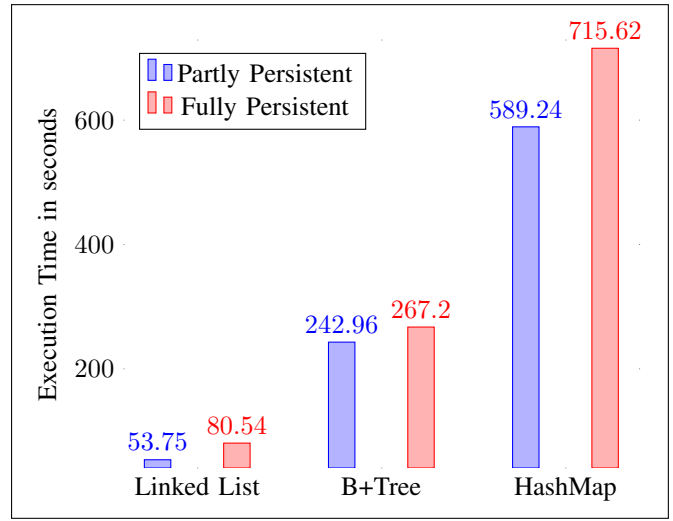

In applications, we find a mix of update operations - interleaving writes and deletes. Hence, we wanted to conduct an experiment that could try emulate that behaviour. This led us to synthesizing workloads with differing mix of Insert/Delete configurations. We check for the following configurations (\texttt{Frequency of Inserts : Frequency of Deletes}) - \texttt{1:1} , \texttt{2:1} and \texttt{4:1} . Figure \ref{fig:1-1}, \ref{fig:2-1} and \ref{fig:4-1} show the results of these experiments. The speedups of the Linked List data structure varies from \texttt{81\%} to \texttt{49\%}. For B+Tree, the speedups are more modest and vary between \texttt{6\%} to \texttt{16\%} depending on the workload. Similarly, we see speedups in the range of \texttt{14\%} to \texttt{21\%} for Hashmaps. \\
In all the above experiments (including Insert-only and Delete-only workloads), we find that the time spent by the program outside of Flush Operations in both Partly Persistent and Fully Persistent stay fairly constant and thus the only reduction is due to the reduction of time in handling flushes.

\begin{figure}[h]
    \centering
\framebox{
\begin{tikzpicture}
  \begin{axis}[
    ybar=20pt,
    x axis line style = { opacity = 0 },
    ylabel            = {Execution Time in seconds},
    tickwidth         = 0pt,
    enlarge x limits  = 0.2,
    enlarge y limits  = 0.02,
    symbolic x coords = {Linked List, B+Tree, HashMap},
    xtick = data,
    nodes near coords,
    legend pos=north west
  ]
\addplot
    coordinates {
(Linked List, 29.873) (B+Tree, 229.843) (HashMap, 463.921)
    };
\addplot
    coordinates {
(Linked List, 54.095) (B+Tree, 267.139) (HashMap, 529.803)
    };
\legend{Partly Persistent, Fully Persistent}

\end{axis}
\end{tikzpicture}
} 
\caption{1:1 (Insert/Delete) Workload : Execution Time}
\label{fig:1-1}
\end{figure}
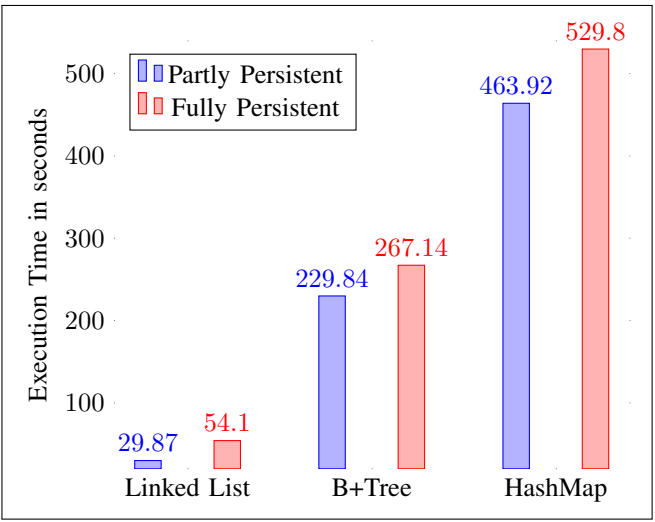

\begin{figure}[h]
    \centering
\framebox{
\begin{tikzpicture}
  \begin{axis}[
    ybar=20pt,
    x axis line style = { opacity = 0 },
    ylabel            = {Execution Time in seconds},
    tickwidth         = 0pt,
    enlarge x limits  = 0.2,
    enlarge y limits  = 0.02,
    symbolic x coords = {Linked List, B+Tree, HashMap},
    xtick = data,
    nodes near coords,
    legend pos=north west
  ]
\addplot
    coordinates {
(Linked List, 39.715) (B+Tree, 243.001) (HashMap, 584.205)
    };
\addplot
    coordinates {
(Linked List, 71.524) (B+Tree, 257.638) (HashMap, 712.421)
    };
\legend{Partly Persistent, Fully Persistent}

\end{axis}
\end{tikzpicture}
} 
\caption{2:1 (Insert/Delete) Workload : Execution Time}
\label{fig:2-1}
\end{figure}

\subsection{Re-flushing the Same Cache Line}
\begin{figure}[h]
    \centering
\framebox{
\begin{tikzpicture}
  \begin{axis}[
    ybar=20pt,
    x axis line style = { opacity = 0 },
    ylabel            = {Execution Time in seconds},
    xlabel            = {Size of Flushed Data},
    tickwidth         = 0pt,
    enlarge x limits  = 0.2,
    enlarge y limits  = 0.02,
    symbolic x coords = {8B, 16B, 32B, 64B},
    xtick = data,
    nodes near coords,
    legend pos=north west
  ]
\addplot
    coordinates {
(8B, 253.180) (16B,244.151) (32B,240.465) (64B,156.929) 
    };

\end{axis}
\end{tikzpicture}
} 
\caption{Re-Flushing Same Cache Line : Execution Time}
\label{fig:diffsizes}
\end{figure}
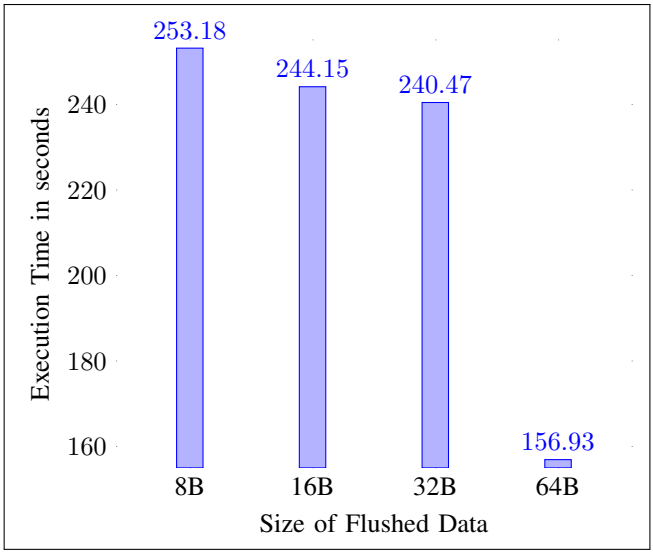

In all our experiments above, the addresses flushed were cache-line aligned (64B) and we collated multiple operations to the same cache-line and flushed them at the same time. That led us to the question, what would be the performance impact if as a software designer we were not careful about the size and alignment of the data we were flushing out. To do that, we created a sample data structure similar to simple Linked List which holds a \texttt{struct} of integer values as its data, and we flush data of different sizes out. We run a micro-benchmark where we INSERT data to this data structure. Figure \ref{fig:diffsizes} shows the result of this experiment. We find that the cost of flushing out non-aligned data ends up causing a huge performance loss, with the 8B data flushes having a slowdown of \texttt{61.3\%} as opposed to the cache-aligned 64B version. This slowdown occurs due to the penalty of re-fetching the same cache-line multiple times. Thus, while designing data structures in Persistent Memory, we pay special attention to the alignment data we are flushing out.  

\subsection{Reconstruction Performance}
To ensure fast performance for the data structure in volatile memory, the redundant data fields are extremely useful. For example, PREV pointer in Linked Lists helps in speeding up traversal in Linked Lists. Hence, there is a need to reconstruct the redundant data fields using the persistently stored data structures, while ensuring that the time taken for reconstruction is lower than the performance gained through storing only a subset of data fields. To ensure reliable reconstruction, we use our second version of the Partly Persistent Implementation. In this version, data structure operates on volatile memory and is mapped to Persistent Memory only on flush operations. 
We find that the time taken for reconstruction varies depending on the complexity of reconstruction. For Hashmaps, it takes 7 seconds to reconstruct a 1.2 GB data structure, whereas it takes 2 seconds for a Linked List data structure of a similar size to be recreated. There is scope for further improvement, for example parallelizing the reconstruction algorithm would make the step much faster.    

\subsection{Correctness}
To ensure correctness of the data structure and our evaluation, we took the following measures:
\begin{itemize}
    \item Added checks to ensure that flush operations only happen for data mapped to Persistent Memory. This ensured that we did not have any stray flushes which might impact our performance results.
    \item We added bugs into our data structure in volatile memory at random locations to check if it impacted our persistent data structure. These bugs were added before a flush, which is the point where we checkpoint our data structure in persistent memory. We were able to correctly recover our data structure.
    \item We added checks to ensure that the data written before a crash was the same as the data recovered through our reconstruction algorithm. 
\end{itemize}
Using these steps we were able to find and fix bugs in our program and we find that all our recoveries work. 

\section{Discussion}
This paper shows that reducing the number of flushes helps in reducing the execution time of the program. But we do not look at the following considerations here:
\begin{itemize}
    \item Fences: The number of fences or more precisely number of flushes per fence also plays a huge rule in determining the program performance. Having multiple flushes per fence allows the overlapping flush latency, and hence reduces the impact of flushes.
    \item Read Overheads: Our work focused on operations that wrote data to cache-lines, and thus we never compared the performance impact of Read operations to Persistent Data Structures. Such a study would require us to compare with an implementation present only in Volatile Memory.
\end{itemize}

We have been basing our evaluation on micro-benchmarks that we ran on these data structures. Running widely-used Key-Value store benchmarks would give us a better idea about the performance improvements that we can expect in real-world applications. 

\section{Conclusion}
With the introduction of Persistent Memory, accesses to Non-Volatile Memory has become cheaper than ever before. But using it as we use normal volatile memory would lead to severe performance degradation due to the higher latency of accessing it. In this paper, we show that by selectively persisting data fields in a data structure, we are able to achieve speedups in the range of 5-20\% and sometimes speedups as high as 165\% for certain workloads.\\
Implementation of the data structures is present at \url{https://github.com/pratyushmahapatra/PersistentApps}

\section{Acknowledgment}
I would sincerely like to thank Prof. Mark D. Hill and Prof. Michael M. Swift for their guidance through my two years of graduate studies. Their insight into the subject is unparalleled and I have always been in awe of them. Getting to work with them help me grow as a person and a researcher. Their advice will be sorely missed as I embark on a new journey in the industry.

\bibliography{references}  
\bibliographystyle{unsrt}

\end{document}